\documentclass[12pt]{iopart}
\usepackage{psfig}
\usepackage{cite}
\usepackage{iopams}

\def\submitto#1{\vspace{28pt plus 10pt minus 18pt}
     \noindent{\small\rm {\it #1}\par}}
\def\abs#1{\left\vert #1 \right\vert}
\def\Zed{\mathbb{Z}}

\begin{document}
\title{Magnetization plateaus in frustrated antiferromagnetic
quantum spin models}

\author{A.\ Honecker\dag\footnote[4]{To
whom correspondence should be addressed ({\tt a.honecker@tu-bs.de})},
J.\ Schulenburg\ddag\ and J.\ Richter\P
}
\address{\dag\ Institut f\"ur Theoretische Physik,
Technische Universit\"at Braunschweig,
38106 Braunschweig, Germany}
\address{\ddag\ Universit\"atsrechenzentrum,
Otto-von-Guericke-Universit\"at Magdeburg,
39016 Magdeburg, Germany}
\address{\P\ Institut f\"ur Theoretische Physik,
Otto-von-Guericke-Universit\"at Magdeburg,
39016 Magdeburg, Germany}


\begin{abstract}
Plateaus can be observed in the zero-temperature magnetization curve
of quantum spin systems at rational values of the magnetization.
In one dimension, the appearance of a plateau is controlled by a quantization
condition for the magnetization which involves the length of the local spin
and the volume of a translational unit cell of the ground state.
We discuss examples of geometrically frustrated quantum spin systems
with large (in general unbounded) periodicities of spontaneous breaking
of translational symmetry in the ground state.

In two dimensions, we discuss the square, triangular and Kagom\'e lattices
using exact diagonalization (ED) for up to $N=40$ sites.
For the spin-$1/2$ $XXZ$ model on the triangular lattice we study
the nature and stability region of a plateau at one third of the saturation
magnetization. The Kagom\'e lattice gives rise to particularly rich
behaviour with several plateaus in the magnetization curve and a jump
due to local magnon excitations just below saturation.

\end{abstract}

\submitto{\JPCM {\bf 16} (2004) S749-S758
 \hfill 
  {\rm version of 30 October 2003}}


\section{Introduction}

Geometrically frustrated quantum spin systems constitute a class of systems
exhibiting interesting quantum phenomena like different (unusual) quantum
phases (see e.g.\ \cite{LhuiMi,RSH} for reviews of two-dimensional models).
An external magnetic field may further enhance this frustration since it
competes with the antiparallel alignment of spins favoured by antiferromagnetic
exchange. This leads to interesting quantum phenomena in the high-field
magnetization process of geometrically frustrated quantum spin systems.
In particular, a spin gap may be opened by the external magnetic field,
giving rise to a plateau in the $T=0$ magnetization curve.
Although no complete review of the subject exists so far, different aspects
are summarized e.g.\ in \cite{LhuiMi,RSH,CGHP,Rice,habil}.
In this article we will illustrate the status of the field by pointing out and
presenting selected recent and new results.

Specifically we consider the $XXZ$ model in an external magnetic field $h$
\begin{equation}
H = \sum_{\langle i,j \rangle} \ J_{i,j} \ \left(
  S^x_i S^x_j + S^y_i S^y_j + \Delta S^z_i S^z_j \right)
 - h \sum_{i} S^z_{i} \, ,
\label{Hxxz}
\end{equation}
where the $\vec{S}_i$ are spin-$S$ operators at site $i$ and
the $J_{i,j}$ are the exchange constants between pairs of sites
$\langle i, j \rangle$ connected in the lattice topology under
consideration. The special case of (\ref{Hxxz}) where the
$XXZ$ anisotropy satisfies $\Delta = 1$ corresponds to the Heisenberg
model. Below we will concentrate on the extreme quantum limit
$S=1/2$ and antiferromagnetic coupling $J_{i,j} \ge 0$ although
some of the results can be generalized e.g.\ to $S > 1/2$.

An important observable is the magnetization
\begin{equation}
\langle M \rangle= {1 \over N\,S} \left\langle
                    \sum_{i} S^z_{i}\right\rangle
\label{defM}
\end{equation}
which we normalize to saturation value $\langle M \rangle=1$
(here $N$ is the total number of spins in the system).
Note that the magnetization (\ref{defM}) is a conserved quantity
for the Hamiltonian (\ref{Hxxz}). This is technically useful
for computing the magnetization curve e.g.\ by exact diagonalization
(ED) using the Lanczos method.

\section{One dimension}

In one dimension,
Oshikawa, Yamanaka and Affleck proposed the following condition on
the magnetization $\langle M \rangle$ at a plateau in the $T=0$
magnetization curve \cite{AOY}
\begin{equation}
S V \left(1 - \langle M \rangle \right) \in \Zed \, .
\label{condM}
\end{equation}
Here $S$ is the size of the local spin and $V$ is the number of spins in a
translational unit cell of the {\it ground state}.
The condition (\ref{condM}) can be understood easily starting from a limit
where the system decouples into clusters with $V$ sites (see e.g.\
\cite{poly}) and in this case $V$ is the number of spins in a
unit cell of the Hamiltonian. However, in particular in frustrated
spin systems one finds that translational symmetry is spontaneously broken
in the ground state (c.f.\ examples in the following sections)
and then $V$ is larger than (namely an integer multiple of) the unit
cell of the Hamiltonian.
Nevertheless, the quantization condition (\ref{condM}) implies that the
magnetization $\langle M \rangle$ always has a rational value on a plateau.

\subsection{Frustrated $S=1/2$ Heisenberg chain}

\begin{figure}[tb]
\centerline{\psfig{figure=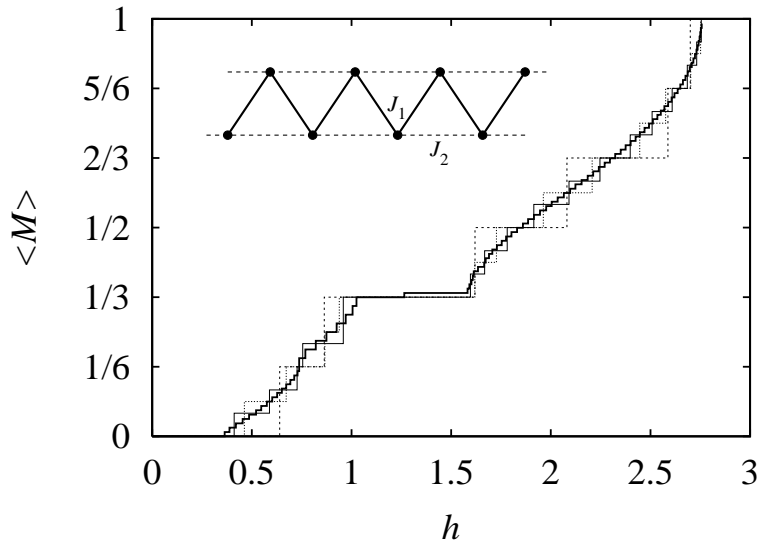,width=0.7\columnwidth}}
\caption{
\label{figZigzag}
Inset:
Heisenberg chain with nearest and next-nearest neighbor coupling
$J_1$ and $J_2$, respectively.
Points denote the locations of $S=1/2$ spins.
Main panel:
Magnetization curve for $J_1 = 1$, $J_2 = 0.8$.
The thin dashed and solid curves were obtained by ED of rings (periodic
boundary conditions) with $L=12$ (dashed), $24$ (dotted) \cite{zigzag}
and $36$ sites (full); the bold solid curve was obtained by DMRG
for $L=192$ sites with open boundary conditions \cite{OkTo}.
}
\end{figure}

\label{secZigzag}

Let us start by discussing some selected aspects of
the $S=1/2$ frustrated Heisenberg chain (see inset of Fig.~\ref{figZigzag}).
Majumdar and Ghosh noticed more than 30 years ago \cite{MaGo1,MaGo2,Maju}
that one can explicitly write down the ground state if 
the next-nearest neighbour coupling $J_2$ and the nearest neighbour
coupling $J_1$ are related by $J_2 = J_1/2$. This so-called
`Majumdar-Ghosh' state is a two-fold degenerate dimer product state and
has become a paradigm for the frustration-induced opening of a spin gap
accompanied by spontaneous breaking of translational symmetry by a period two. 

In the presence of a magnetic field, the spin gap leads to
an $\langle M \rangle = 0$ plateau in the magnetization curve.
Starting with \cite{ToHa}, many authors have studied the magnetization process
of this model by ED (see e.g.\ \cite{zigzag} and references
therein) and no plateaus with $\langle M \rangle \ne 0$ were found.
Finally, a recent density-matrix renormalization group (DMRG)
study of the $S=1/2$ frustrated Heisenberg chain exhibited a clear
plateau at $\langle M \rangle = 1/3$ in the region of strong frustration
$J_1 \approx J_2$ \cite{OkTo}, as is illustrated by the main panel
of Fig.~\ref{figZigzag} which reproduces the magnetization curve
for $J_2 = 0.8\, J_1$
(the step on the $\langle M \rangle = 1/3$ plateau in the
DMRG curve is due to the open boundaries).
The state of this $\langle M \rangle = 1/3$
plateau spontaneously breaks translational symmetry by a period {\it three}.
Two points are worthwhile noticing here. First, this period three is higher
than the period two appearing in the Majumdar-Ghosh state
\cite{MaGo1,MaGo2,Maju} at $h=0$.
Second, the very presence of this plateau was missed for a long time.

With hindsight, this $\langle M \rangle = 1/3$ plateau can be observed
also in the ED results for $L=12$ and $24$ sites and periodic boundary
conditions (reproduced for $J_2 = 0.8\, J_1$ in Fig.~\ref{figZigzag}
after \cite{zigzag} -- this figure also includes a new $L=36$ curve). It may
be instructive to consider the reasons for missing it nevertheless.
First, the excited states at $h=0$ \cite{WhAff} and correspondingly the
ground states in a magnetic field $h\ne 0$ have in general incommensurate
momenta \cite{zigzag} which leads to oscillating
finite-size effects. Second, one should concentrate on system sizes that
can accommodate the known antiferromagnetic ground states in the limit
of two decoupled chains $J_1 \to 0$, implying
that $L$ should be a multiple of 4. In addition, only those values of
the magnetization can be studied which are realized for a given
system size. In particular, $\langle M \rangle = 1/3$ is realized only
if $L$ is a multiple of 3. These considerations restrict the system
sizes that should be used for the discussion of an $\langle M \rangle = 1/3$
plateau in the frustrated Heisenberg chain to multiples of 12. The first
cases are $L=12$, $24$ and $36$. These are used in Fig.~\ref{figZigzag}
and are in good agreement with the DMRG results of \cite{OkTo}.

\subsection{Orthogonal dimer chain}

\begin{figure}[tb]
\centerline{\psfig{figure=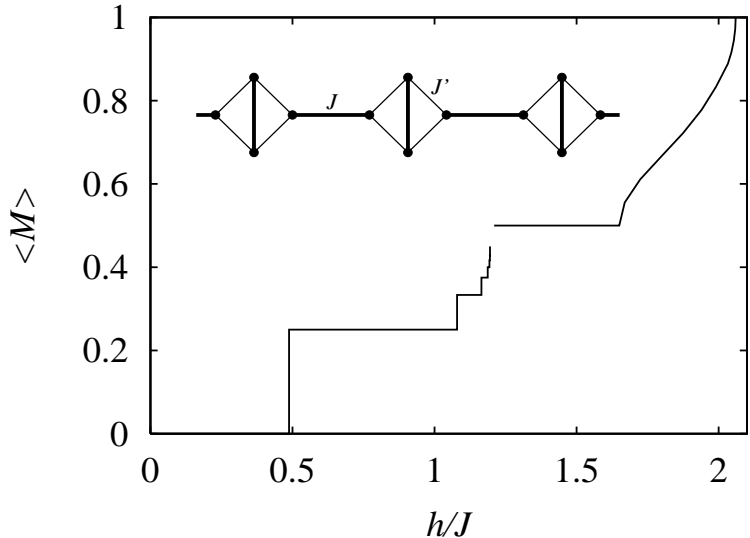,width=0.7\columnwidth}}
\caption{
\label{figOrthDim}
Inset: Orthogonal dimer chain.
Points denote the locations of $S=1/2$ spins.
Main panel:
Magnetization curve for $J' = 0.7 J$ \cite{SchuRi}.
Note the long (probably infinite) sequence of plateaus for
$\langle M \rangle < 1/2$.}
\end{figure}

Now we wish to draw the reader's attention to the $S=1/2$ orthogonal dimer
chain (also known as dimer-plaquette chain) \cite{IvRi,RIS} shown in the
inset of Fig.~\ref{figOrthDim}, a further one-dimensional model where some
remarkable properties were missed originally.
This model attracted renewed attention
recently since it can be regarded as a one-dimensional version
of the two-dimensional Shastry-Sutherland model \cite{ShaSu}.

A first study \cite{KOK} of the magnetization process of the $S=1/2$
orthogonal dimer chain found plateaus with $\langle M \rangle = 1/4$
and $1/2$. However, this study missed further plateaus and
thus the remarkable property of the orthogonal dimer chain that
it gives rise to a magnetization curve with (probably infinitely) many
plateaus for some values of the parameters
\cite{SchuRi}. At least some steps in the derivation can be carried
out exactly exploiting the special property of the orthogonal dimer chain
that the total spin on each vertical dimer is conserved.
A plateau-state can then be characterized as follows \cite{SchuRi}:
$k$ consecutive vertical dimers form triplets ($S=1$),
separated by singlets ($S=0$). Covering the chain periodically
with such `fragments' of length $k$ each in the sector $S^z = k$
yields
\begin{equation}
\langle M \rangle = {k \over 2k + 2} \, .
\label{Mfrag}
\end{equation} 
At $J' = 0.7 J > 0$ and for $\langle M \rangle < 1/2$
only precisely these states consisting
of fragments of length $k$ with $S^z = k$ appear as ground states
in a magnetic field \cite{SchuRi}, whereas for $\langle M \rangle > 1/2$
the ground state is in a sector where all vertical dimers form triplets
$S = 1$ (one difference between the present Fig.~\ref{figOrthDim}
and the original Fig.~3 of \cite{SchuRi} is in the system sizes that
have been used for extrapolating the magnetization curve in the region
$\langle M \rangle \ge 1/2$).
As long as $\langle M \rangle < 1/2$,
fragments of increasing length become ground states
with increasing field, leading to a magnetization curve with
infinitely many plateaus at the magnetization values given by (\ref{Mfrag})
and jumps in between, compare the main panel of Fig.~\ref{figOrthDim}.
This infinite series accumulates at $\langle M \rangle = 1/2$, where one
finds a further pronounced plateau before a smooth transition to saturation
follows.

The infinite sequence of plateaus arises in the
orthogonal dimer chain because the ground states
manifestly break translational symmetry with arbitrarily long periodicity
$k+1$, implying that at least in one dimension there is no general upper
bound on the possible periods for spontaneous breaking of translational
symmetry.

\section{Two dimensions}

\label{sec2D}

To the best of our knowledge there are no rigorous arguments why
the condition (\ref{condM}) should hold in dimensions higher than one.
Furthermore, starting from two dimensions one may have true long-range
order in the spin components at $T=0$. Accordingly, an external magnetic
field can induce transitions between different ordered and disordered states.

In one dimension, a plateau at a given
$\langle M \rangle$ already requires a detailed analysis of the
specific model. And, even more so the presence and nature of a plateau state
must be discussed case by case in two dimensions.
Some selected examples are presented in the following sections.

\begin{figure}[tb]
\centerline{\psfig{figure=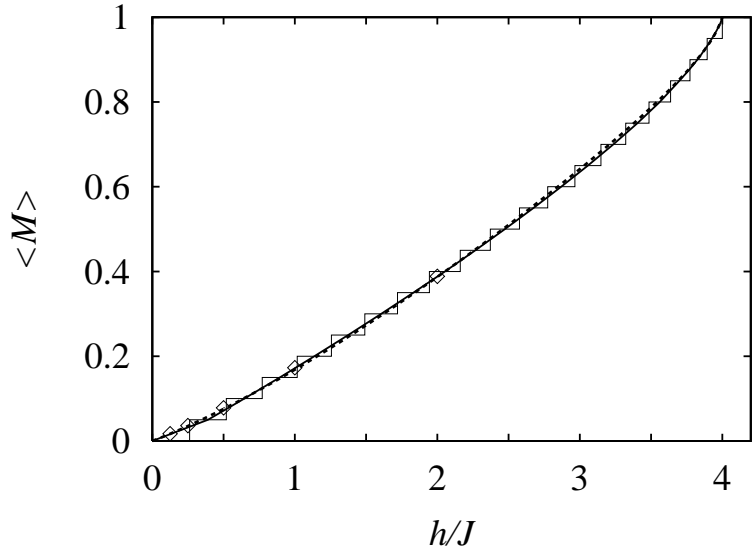,width=0.7\columnwidth,angle=270}}
\caption{
\label{figSquare}
Magnetization curve of the $S=1/2$ Heisenberg antiferromagnet on the
square lattice. The thin solid line is for $N=40$ sites,
the full bold line is an extrapolation to the thermodynamic limit.
Also shown are a second-order spinwave result \cite{ZhNi} (dashed bold line)
and QMC results \cite{sandvik} (diamonds).}
\end{figure}

\subsection{Square lattice}

\label{secSquare}

Let us start with a brief discussion of the magnetization process of the square
lattice which is a non-frustrated model and well understood.
Fig.~\ref{figSquare} shows the magnetization curve of the
$S=1/2$ square-lattice Heisenberg model ($\Delta = 1$) obtained by
different approaches. First, the thin full line shows the result obtained
by ED for a finite lattice with $N=40$ sites (see also \cite{YaMue,comp2D} for
earlier ED studies -- in the particular case $N=40$,
the largest dimension to be treated was 554 596 326 and occurred
in the sector $S^z = 3$). The full bold
line denotes an extrapolation of the ED data to the thermodynamic limit which
is obtained by connecting the midpoints of the finite-size steps
at the largest available system size. One observes a smooth magnetization
curve with no peculiar features (in particular no plateaus) for
$\abs{\langle M \rangle} < 1$.
Note that close to saturation the extrapolated curve includes data at
large system sizes, which is not shown explicitly in Fig.~\ref{figSquare}.
More precisely, for
$\langle M \rangle \ge 0.84375$, the curve is based exclusively on
finite lattices with at least $8 \times 8$ sites. The high-field part of the
magnetization curve is therefore particularly well controlled by ED.

Second, a second-order spinwave result \cite{ZhNi} is also available
and shown by the bold dashed line in Fig.~\ref{figSquare}. Third,
the magnetization process of the square lattice can also be studied
by quantum Monte Carlo (QMC) since this lattice is not frustrated.
The diamonds in Fig.~\ref{figSquare} show
available stochastic-series-expansion QMC results \cite{sandvik}.

The quantitive differences of the results of all three approaches
are small, i.e.\ each approach yields a good description of
the $S=1/2$ square lattice. Since the spinwave approach \cite{ZhNi} is based
on a N\'eel state, we may therefore conclude that N\'eel order prevails in
the transverse components for $\abs{\langle M \rangle} < 1$
(see also \cite{YaMue} for a discussion from the point of view of ED).

\subsection{Triangular lattice}

\label{secTriag}

The $S=1/2$ $XXZ$ model on the triangular
lattice is among the first models whose magnetization process was
studied by ED \cite{NiMi}. These early studies already found a plateau
with $\langle M \rangle = 1/3$, at least for Ising-like anisotropies
$\Delta > 1$. Due to the restriction to at most 21 sites, it was first not
completely clear whether the plateau persists in the
isotropic regime $\Delta \approx 1$.
The magnetization process of the isotropic model was analyzed
further using spinwave theory \cite{ChuGo}. This study
provided evidence that the $\langle M \rangle = 1/3$ plateau exists
also at $\Delta = 1$ and estimates for its boundaries were obtained.

\begin{figure}[tb]
\centerline{\psfig{figure=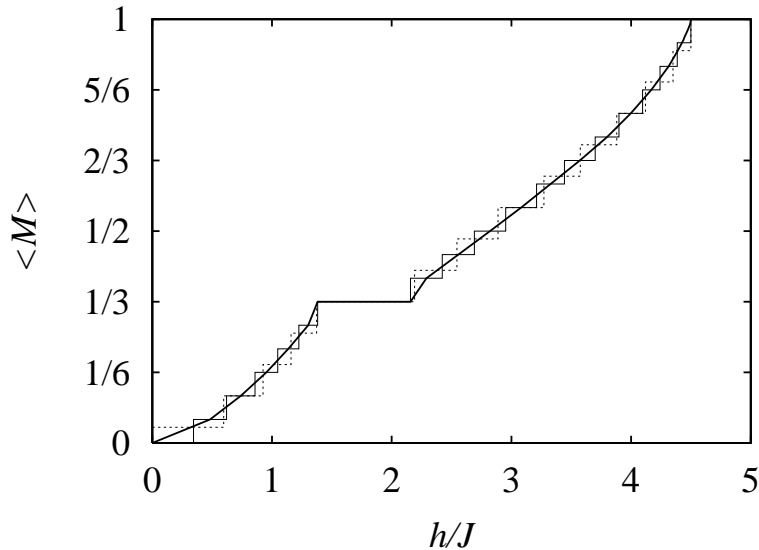,width=0.7\columnwidth,angle=270}}
\caption{
\label{figTriag}
Magnetization curve of the $S=1/2$ Heisenberg antiferromagnet on the
triangular lattice.
The thin dashed and solid line are for $N=27$ and $N=36$ sites, respectively.
The bold line is an extrapolation to the thermodynamic limit.
}
\end{figure}

Fig.~\ref{figTriag} shows the magnetization curves obtained by
ED for $\Delta = 1$ on finite lattices with $N=27$ and $36$ sites (thin lines).
Both curves exhibit a clear plateau at $\langle M \rangle = 1/3$ in an
otherwise smooth magnetization curve.
These two curves overlap with previous ED results \cite{comp2D,BLLP}.
Note, however, that for $N=36$ only the sector
$\vec{k}=0$ was studied previously for $\langle M \rangle \le 1/3$ \cite{BLLP},
but it turns out that the ground state is not always in this sector.
Inclusion of sectors with $\vec{k} \ne 0$ for $N=36$, a different
shape of the $N=27$ system and a previously incomplete $N=36$
curve lead to small differences of our Fig.~\ref{figTriag} as compared
to Fig.~4 of \cite{comp2D}.
The first-order spinwave results for the magnetic fields
at the lower and the upper boundaries of the $\langle M \rangle = 1/3$
plateau \cite{ChuGo} are smaller by about $0.13 J$ (lower boundary)
and $0.01 J$ (upper boundary) than the ED results presented here for
$N=36$ and $S=1/2$.

The full bold line in Fig.~\ref{figTriag} denotes an extrapolation of
the ED data to the thermodynamic limit which is obtained by connecting
the midpoints of the finite-size steps at the largest available system size
(except for the boundaries of the $\langle M \rangle = 1/3$ plateau where
corners were used). Close to saturation this includes again bigger
system sizes than those explicitly shown in Fig.~\ref{figTriag}.

\begin{figure}[tb]
\centerline{\psfig{figure=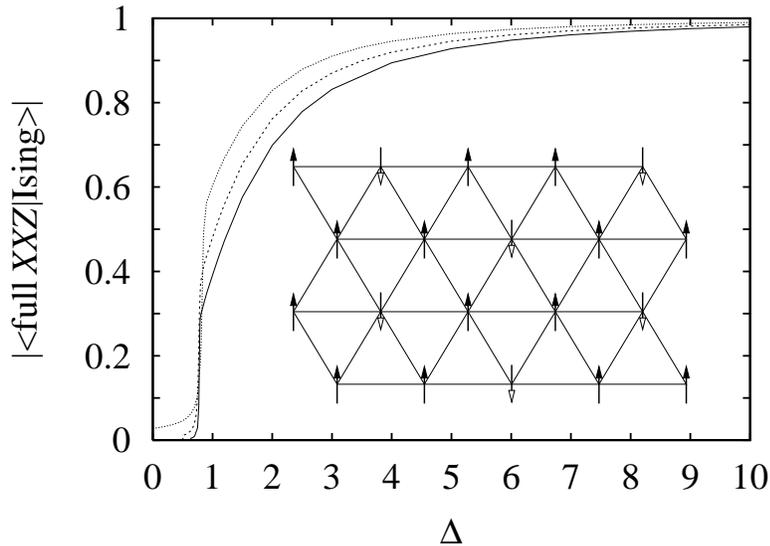,width=0.7\columnwidth}}
\caption{
\label{figTrOv}
Overlap of the wave function $\vert\hbox{full }XXZ\rangle$
of the full $S=1/2$ $XXZ$ model on the triangular lattice at
$\langle M \rangle = 1/3$ with the corresponding state
$\vert\hbox{Ising}\rangle$ of the Ising model
as a function of the $XXZ$ anisotropy $\Delta$.
Lines are for finite lattices with $N=18$ (dotted), $27$ (dashed) and
$36$ (full) sites. The inset shows one representative
state $\vert\hbox{Ising}\rangle$ for the three degenerate
ground states of the Ising model at $\langle M \rangle = 1/3$.
}
\end{figure}

The state of the $\langle M \rangle = 1/3$ plateau can be easily
understood in the Ising limit $\Delta \gg 1$ \cite{miyashita,comp2D}.
Quantum fluctuations are completely suppressed in the limit $\Delta \to \infty$
and the $\langle M \rangle = 1/3$ state is a classical state
where all spins on two of the three sublattices of the triangular lattice
point up and all spins on the third sublattice point down, as is
sketched in the inset of Fig.~\ref{figTrOv}. This state corresponds
to an ordered collinear spin configuration. It breaks translational symmetry
by a period three and accordingly is threefold degenerate.
One can then use perturbation theory in $1/\Delta$ to study the
$\langle M \rangle = 1/3$ plateau of the $XXZ$ model \cite{comp2D}.
Here we numerically compute the overlap of the appropriate linear
combination $\vert\hbox{Ising}\rangle$ of the three Ising states
and the $\langle M \rangle = 1/3$ wave function $\vert\hbox{full }XXZ\rangle$
of the full $XXZ$ model with $S=1/2$ in order to show that this description
remains qualitatively valid even in the isotropic region $\Delta \approx 1$.
Fig.~\ref{figTrOv} shows results for the overlap
$\abs{\langle \hbox{full }XXZ \vert \hbox{Ising}\rangle}$ on finite lattices
with $N=18$, $27$ and $36$ sites
(for $N=36$ we have computed the overlaps by considering
only the wave function within the $\vec{k}$-subspace that
contains the ground state for $\Delta \ge 1$).
One observes that this overlap remains
large even in the vicinity of the Heisenberg model $\Delta = 1$ and
drops sharply around $\Delta = 0.75$. Accordingly, this
$\langle M \rangle = 1/3$ plateau state is a stable phase for
$\Delta > \Delta_c$ with $\Delta_c \approx 0.75$. Remarkably, this
estimate for $\Delta_c$
is close to the estimate $\Delta_c \approx 0.85$ for the vanishing point of
the $\langle M \rangle = 1/3$ plateau which was obtained in
\cite{comp2D} by comparing the plateau widths on $3 \times 6$ and
$3 \times 9$ systems. Note that the location of the sharp drop in
Fig.~\ref{figTrOv} is almost independent of system size although the
absolute values of the overlaps
$\abs{\langle \hbox{full }XXZ \vert \hbox{Ising}\rangle}$ do depend on $N$.
We can thus locate the point $\Delta_c$ where
the $\langle M \rangle = 1/3$ plateau disappears quite accurately
at $\Delta_c = 0.76 \pm 0.03$.

\subsection{Kagom\'e lattice}

\label{secKag}

\begin{figure}[tb]
\centerline{\psfig{figure=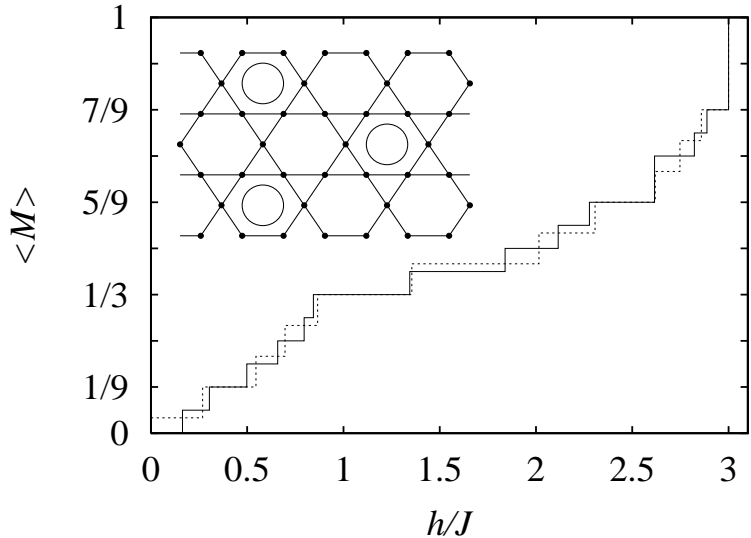,width=0.7\columnwidth}}
\caption{
\label{figKag}
Magnetization curve of the $S=1/2$ Heisenberg antiferromagnet on the
Kagom\'e lattice with $N=27$ (dashed line) and $36$ sites (solid line).
The inset shows part of the Kagom\'e lattice with a
$\sqrt{3} \times \sqrt{3}$ superstructure indicated by the
circles in certain hexagons.
}
\end{figure}

Finally, Fig.\ \ref{figKag} shows magnetization curves
for the Kagom\'e lattice (see inset) with $N=27$
\cite{hida,jump} and $36$ sites. Note that the $N=36$ curve in
Fig.\ \ref{figKag} differs slightly from the corresponding
curve in \cite{jump} for $\langle M \rangle \le 1/3$. The reason
is that due to the computational effort only selected $\vec{k}$-sectors
were investigated in \cite{jump} and in some cases the correct symmetry of the
ground state was in fact missed. Our new $N=36$ curve agrees
with unpublished results of Waldtmann and Everts \cite{WEpriv}.

The Kagom\'e lattice is famous for a disordered ground state
at $h=0$ accompanied by a small spin gap
with many singlets inside this gap (see \cite{WEBLSLP,mila}
and references therein). The spin gap should give rise to
a narrow $\langle M \rangle = 0$ plateau even if it is
difficult to recognize in Fig.\ \ref{figKag}.

A plateau at $\langle M \rangle = 1/3$ can be recognized better
in Fig.\ \ref{figKag}. In fact, the presence of this plateau
at $\langle M \rangle = 1/3$ in the $S=1/2$ Heisenberg model on
the Kagom\'e lattice has been established previously by considering
also system sizes different from those shown in Fig.~\ref{figKag}
\cite{hida,CGHP02}. Nevertheless, the state of this plateau is
still an interesting issue.
For the classical Heisenberg model at $\langle M \rangle = 1/3$,
thermal fluctuations select collinear states, but due to
the huge degeneracy of this space, there appears to be no real order
on the classical level at $\langle M \rangle = 1/3$ \cite{zhitomirsky}
(see also \cite{ShHo}).
For $S=1/2$, it is useful to consider the $XXZ$ model. In the
Ising limit $\Delta \to \infty$ one can then first establish \cite{CGHHPRS}
a relation to a quantum dimer model on the
hexagonal lattice which was argued \cite{MS2001,MSC2001} to give
rise to a valence bond crystal ground state with a
$\sqrt{3} \times \sqrt{3}$ order indicated by the circles in the
hexagons in the inset of Fig.\ \ref{figKag}.
Furthermore, like for the triangular lattice
one can compute the overlap (see Fig.~\ref{figTrOv})
of the $\langle M \rangle = 1/3$ wave function of the $XXZ$ model
with that of the quantum dimer model as a function of $\Delta$
and one finds no evidence for a phase transition for $\Delta \ge 1$
\cite{CGHHPRS}. This implies that also the $\langle M \rangle = 1/3$
state of the $S=1/2$ Heisenberg model on the Kagom\'e lattice is a
three-fold degenerate valence bond crystal, i.e.\ an ordered state with
many low-lying non-magnetic excitations, but still a small gap to
{\it all} excitations \cite{CGHHPRS}.

There may be a further plateau at $\langle M \rangle = 5/9$ in
Fig.\ \ref{figKag} although it is difficult to draw unambiguous conclusions
from the available numerical data in this region of magnetization values.

Finally, one can see a pronounced jump of height
$\delta \langle M \rangle = 2/9$
just below saturation in Fig.\ \ref{figKag}. This jump arises due to
local magnon excitations which can be constructed in the spin-$S$ $XXZ$ model
on the Kagom\'e lattice \cite{jump,RSHSS}.
These local magnon excitations give rise to an exact degeneracy at the
saturation field and thus a macroscopic jump in the magnetization curve
of height $\delta \langle M \rangle = 1/(9 S)$.
In the state just below the jump, magnons are localized on the
hexagons marked by circles in the inset of Fig.\ \ref{figKag}.
Since this is a crystalline ordered state, general arguments
(see e.g.\ \cite{MoTo,oshikawa}) predict a gap
and consequently a plateau in the magnetization curve at
$\langle M \rangle = 1-1/(9 S)$.
Indeed, a plateau at $\langle M \rangle = 7/9$ can be seen clearly
in the magnetization curve
of the $S=1/2$ Heisenberg model on the Kagom\'e lattice in particular
when one considers also lattices with $N=45$ and $54$ sites \cite{jump}.

\section{Conclusions and outlook}

In this article we have discussed examples of plateaus in one- and
two-dimensional quantum spin models. In one dimension, the appearance
of a plateau is controlled by the quantization condition (\ref{condM})
which implies in particular that the magnetization $\langle M \rangle$
on a plateau must be rational. This condition can be interpreted as a
commensurability condition for the ground state. Hence, it is
important to observe that translational symmetry can be spontaneously broken
in frustrated quantum spin models. Even in well-studied examples such as
the frustrated $S=1/2$ Heisenberg chain, an $\langle M \rangle=1/3$ plateau
with a spontaneously broken period three was discovered only recently
\cite{OkTo}. Furthermore, an infinite sequence of plateaus in the orthogonal
dimer chain \cite{SchuRi} shows that there is no upper bound on the
possible periodicity of spontaneous breaking of translational symmetry.

It is less clear if the condition (\ref{condM}) is
also applicable to two and higher dimensions. Nevertheless, it works
for most examples discussed here: The state of the $\langle M \rangle = 1/3$
plateau in the $S=1/2$ triangular lattice has a unit cell with $V=3$ spins
(see section \ref{secTriag}) and for the Kagom\'e lattice the plateaus with
$\langle M \rangle = 1/3$ in the $S=1/2$ model as well
as the one with $\langle M \rangle = 1-1/(9 S)$ in the spin-$S$ model
both have unit cells with $V=9$ spins
(see section \ref{secKag}, in particular the inset of Fig.\ \ref{figKag}).
These cases fit well with the quantization condition (\ref{condM}).
However, the disordered ground state at $h=0$ in the $S=1/2$ Kagom\'e lattice
(see \cite{WEBLSLP,mila} and references therein) is not covered
by the condition (\ref{condM}) since it gives rise to an $\langle M \rangle = 0$
plateau without the appropriate breaking of translational symmetry. 

{}From the point of view of experiments, the exchange constants $J_{i,j}$
dictate the scale of magnetic fields needed to access the high-field
region. SrCu$_2$(BO$_3$)$_2$ is
an $S=1/2$ system with sufficiently small $J$'s that permit the observation
of plateaus at $\langle M \rangle = 1/8$, $1/4$ and $1/3$
in pulsed-field magnetization experiments \cite{KYSMOKKSGU,OKNKUG}.
SrCu$_2$(BO$_3$)$_2$ is believed to be a good realization
of the two-dimensional Shastry-Sutherland lattice \cite{ShaSu}.
Because of its relation to SrCu$_2$(BO$_3$)$_2$, this frustrated spin
model has been analyzed theoretically in detail (see \cite{MiUe}
for a recent review).
We hope that the examples discussed in this article will stimulate further
theoretical and experimental research on the magnetization process
of frustrated quantum spin models.

\ack

We would like to thank D.C.\ Cabra, M.D.\ Grynberg, P.C.W.\ Holdsworth
and P.\ Pujol for useful discussions.
The more complicated computations presented in this article have
been performed on the compute-servers {\tt cfgauss} and
{\tt wildfire} at the computing centers of the TU Braunschweig
and Magdeburg University, respectively. We are particularly grateful
to J.\ Sch\"ule of the Rechenzentrum at the TU Braunschweig
for technical support.
This work was partly supported by the DFG (project Ri615/10-1).

\end{document}